\documentclass[usenatbib]{mn2e}
\usepackage{natbibmnfix,graphicx,times}

\bibpunct{(}{)}{;}{a}{}{,}
\defcitealias{BKH05}{BKH}

\newcommand{\apj}{ApJ}
\newcommand{\apjl}{ApJ}

\newcommand{\aj}{AJ}
\newcommand{\mnras}{MNRAS}
\newcommand{\physrep}{Physics Reports}

\newcommand{\sm}{\mbox{  }{\rm M_\odot}}
\newcommand{\scriN}{{\cal N}}
\newcommand{\delcoll}{\delta_{\mathrm{coll}}}
\newcommand{\lcdm}{$\Lambda$CDM }
\newcommand{\Om}{\Omega_{\rm M}}

\newcommand{\lsim}{\mathrel{\hbox{\rlap{\lower.55ex\hbox{$\sim$}} \kern-.3em \raise.4ex \hbox{$<$}}}}
\newcommand{\gsim}{\mathrel{\hbox{\rlap{\lower.55ex\hbox{$\sim$}} \kern-.3em \raise.4ex \hbox{$>$}}}}

\title{Supermassive Black Hole Merger Rates: Uncertainties from Halo Merger Theory}
\author[Erickcek, Kamionkowski and Benson]{Adrienne L. Erickcek$^1$\thanks{E-mail: erickcek@tapir.caltech.edu (ALE); kamion@tapir.caltech.edu (MK); abenson@tapir.caltech.edu (AJB)}, Marc Kamionkowski$^1$ and Andrew J. Benson$^2$\\
$^1$Division of Physics, Mathematics, \& Astronomy, California Institute of Technology, Mail Code 130-33, Pasadena, CA 91125, USA\\
$^2$Department of Physics, University of Oxford, Keble Road, Oxford OX1 3RH, UK}

\begin{document}
\maketitle
\begin{abstract}
The merger of two supermassive black holes is expected to produce a gravitational-wave signal detectable by the satellite LISA.  The rate of supermassive-black-hole mergers is intimately connected to the halo merger rate, and the extended Press--Schechter formalism is often employed when calculating the rate at which these events will be observed by LISA.  This merger theory is flawed and provides two rates for the merging of the same pair of haloes.  We show that the two predictions for the LISA supermassive-black-hole-merger event rate from extended Press--Schechter merger theory are nearly equal because mergers between haloes of similar masses dominate the event rate.  An alternative merger rate may be obtained by inverting the Smoluchowski coagulation equation to find the merger rate that preserves the Press--Schechter halo abundance, but these rates are only available for power-law power spectra.  We compare the LISA event rates derived from the extended Press--Schechter merger formalism to those derived from the merger rates obtained from the coagulation equation and find that the extended Press--Schechter LISA event rates are thirty percent higher for a power spectrum spectral index that approximates the full \lcdm result of the extended Press--Schechter theory.    
\end{abstract}

\begin{keywords}
black hole physics--gravitational waves--galaxies:haloes--cosmology:theory
\end{keywords}

\section{Introduction}
Structure formation proceeds hierarchically, with small over-dense regions collapsing to form the first dark-matter haloes.  These haloes then merge to form larger bound objects.  The extended Press--Schechter (EPS) formalism  provides a description of ``bottom-up'' structure formation by combining the Press--Schechter halo mass function \citep{PS74} with the halo merger rates derived by \citet{LC93}.  Since its inception, the EPS theory has been an invaluable tool and has been applied to a wide variety of topics in structure formation \citep*[see][and references therein]{BKH05}.

Unfortunately, the Lacey--Cole merger-rate formula, which is the cornerstone of EPS merger theory, is mathematically inconsistent \citep{BKH05}.  It is possible to obtain \emph{two} equally valid merger rates for the same pair of haloes from the EPS formalism.  These two merger rates are nearly equal when the masses of the two haloes differ by less than a factor of one hundred, but they diverge rapidly for mergers between haloes with larger mass ratios.  Consequently, any application of EPS merger theory gives two answers, and if the calculation involves mergers between haloes of unequal masses, the discrepancy between these two predictions may be large.  

Motivated by the ambiguity in the Lacey--Cole merger rate, \citet*[hereafter BKH]{BKH05}, proposed a method to obtain self-consistent halo merger rates.  Since haloes are created and destroyed through mergers, the halo merger rate determines the rate of change of the number density of haloes of a given mass.  By inverting the Smoluchowski coagulation equation \citep{Smol16}, \citetalias{BKH05} find merger rates that predict the same halo population evolution as the time derivative of the Press--Schechter mass function.  In addition to eliminating the flaw that resulted in the double-valued rates in EPS theory, the BKH merger rates by definition preserve the Press--Schechter halo mass distribution when used to evolve a population of haloes.  The Lacey--Cole merger rate fails this consistency test as well, and the use of EPS merger trees has been constrained by this inconsistency \citep*[e.g.][]{MHN01}.

There are three limitations to the BKH merger rates.  First, they are not uniquely determined because the Smoluchowski equation does not provide sufficient constraints on the merger rate.  The BKH merger rate is the smoothest, non-negative function that satisfies the coagulation equation; it exemplifies the properties of a self-consistent merger theory, but it is not a definitive result.  Second, the inversion of the Smoluchowski equation is numerically challenging and solutions have been obtained only for power-law density power spectra.  Finally, the BKH merger rates are derived from the Press--Schechter halo mass function rather than the mass functions obtained from N-body simulations \citep{ST99,Jenkins01}. 

In this paper, we explore the possible quantitative consequences of our limited understanding of merger rates for one of the astrophysical applications of merger theory: the merger rate of supermassive black holes.  Since supermassive black holes (SMBHs) are believed to lie in the centre of all dark-matter haloes above some critical mass, halo mergers and SMBH mergers are intimately related.  By considering only halo mergers that would result in a SMBH merger, the EPS merger rates have been used to obtain SMBH merger rates \citep{Haeh94,MHN01,WL03gw,SHMV04,SHMV05,RW05}. 

SMBH mergers are of great interest because they produce a gravitational-wave signal that may be detectable by the Laser Interferometry Space Antenna (LISA), which is scheduled for launch in the upcoming decade.  Consequently, EPS merger theory has been used to obtain estimates for the SMBH merger event rate for LISA \citep{Haeh94,MHN01,WL03gw,SHMV04,SHMV05,RW05}.  In addition to their intrinsic interest as a probe of general relativity, there is hope that LISA's observations of SMBH mergers would provide a new window into astrophysics at high redshifts.  \citet{WL03gw} used EPS merger theory to derive a redshift-dependent mass function for haloes containing supermassive black holes and then used EPS merger theory to predict the LISA event rate that arises from this SMBH population.  Since SMBH formation becomes more difficult after reionization due to the limitations on cooling imposed by a hot intergalactic medium, the Wyithe--Loeb SMBH mass function and corresponding LISA event rate are highly sensitive to the redshift of reionization.  \citet{MHN01} used EPS merger trees to demonstrate that LISA observes more SMBH merger events when SMBHs at redshift $z=5$ are only found in the most massive haloes as opposed to being randomly distributed among haloes. \citet{KZ06} also used EPS merger trees to show that higher-mass seed black holes ($M_{\rm BH}\sim 10^5 \sm $ as opposed to $M_{\rm BH}\sim 10^2 \sm $) at high redshifts result in significantly higher LISA SMBH-merger event rates.  Unfortunately, these ambitions of using  LISA SMBH-merger event rates to learn about reionization and SMBH formation rest on the shaky foundation of extended Press--Schechter merger theory.  

We first review how the rate of mergers per comoving volume translates to an observed event rate in a \lcdm universe and how the mass of the halo is related to the mass of the SMBH at its centre in Sections \ref{sec:event_rates} and \ref{sec:BHmass}.  In Section \ref{sec:EPS}, we use the EPS formalism to derive an event rate for LISA.  Throughout the calculation, we present the results derived from \emph{both} versions of the Lacey--Cole merger rate.  In Section \ref{sec:BKH}, we explore the alternative merger-rate formalism proposed by \citet*{BKH05}.  Since the BKH merger rates are only available for power-law density power spectra, it is not possible to use them to make a new prediction of the SMBH merger rate and the corresponding event rate for LISA.  Instead, in Section \ref{sec:BKHvsEPS}, we use the event rates for power-law power spectra derived from the EPS and BKH merger theories to gauge how the LISA event rates may be affected by switching merger formalisms. Finally, in Section \ref{sec:CON}, we summarize our results and discuss how these ambiguities in halo merger theory limit our ability to learn about reionization and supermassive-black-hole formation from LISA's observations. 
 
\section{Cosmological event rates}
\label{sec:event_rates}
The merger of two supermassive black holes will produce a gravitational-wave burst.  The observed burst event rate depends on the number density and frequency of black-hole mergers: the number of observed gravitational-wave bursts per unit time ($B$) that originate from a shell of comoving radius $R(z)$ and width ${\rm d}R$ is
\begin{equation}
{\rm d}B=(1+z)^{-1} \scriN(z) \mbox{  }4\pi R^2 \mbox{  }{\rm d}R, 
\label{dBdR}
\end{equation}
where $\scriN(z)$ is the SMBH merger rate per comoving volume as a function of redshift.  The factor of $(1+z)^{-1}$ in equation (\ref{dBdR}) results from cosmological time dilation.  In equation (\ref{dBdR}), and throughout this article, we assume a flat $\Lambda$CDM universe.  Given the relation between comoving distance and redshift, ${\rm d}R = [c/H(z)]\mbox{ }{\rm d}z$, equation (\ref{dBdR}) may be converted to a differential event rate per redshift interval,
\begin{equation}
\frac{{\rm d}B}{{\rm d}z} = (1+z)^{-1} \left(\frac{4\pi [R(z)]^2\scriN(z)c}{H_0\sqrt{\Om(1+z)^3+\Omega_\Lambda}}\right),
\label{dBdz}
\end{equation}
where $\Om$ and $\Omega_\Lambda$ are the matter and dark-energy densities today in units of the critical density.  

The observed gravitational-wave burst rate from SMBH mergers is obtained by integrating equation (\ref{dBdz}) over the redshifts from which the bursts are detectable.  LISA will be able to detect nearly all mergers of two black holes with masses greater than $10^4 \sm $ and less than $10^8 \sm $ up to $z\lsim9$ \citep{Haeh94, SHMV05, RW05}.  Since more massive binary-black-hole systems emit gravitational radiation at lower frequencies and the observed frequency decreases with redshift, very distant ($z\sim9$) mergers of SMBHs with masses greater than $10^8 \sm $ produce signals below LISA's frequency window \citep{SHMV05,RW05}.  However, the number density of $10^8 \sm $ haloes is exponentially suppressed at redshifts greater than four, so it is extremely unlikely that two black holes larger than $10^8 \sm $ will merge at redshifts $z\gsim4$.  Thus, the upper bounds on the relevant redshift and SMBH mass intervals are determined by the population of supermassive black holes and not LISA's sensitivity.  

\section{The relationship between halo mass and black hole mass}
\label{sec:BHmass}

The transition from the rate of halo mergers to the rate of detectable SMBH mergers [$\scriN(z)$ as defined in equation (\ref{dBdR})] requires a relationship between the mass of a halo and the mass of the SMBH at its centre.  Since LISA is sensitive to SMBH mergers at high redshifts, this $M_{\rm BH}-M_{\rm halo}$ relation must be applicable to high redshifts as well.

Observations of galaxies out to $z\sim3$ reveal a redshift-independent correlation between the mass of the central black hole and the bulge velocity dispersion $\sigma_{\rm c}$ \citep{FM00,GRKetal00, Tremaine02}.  The connection between $\sigma_{\rm c}$ and halo mass is mediated by the circular velocity $v_{\rm c}$.  Using a sample of thirteen spiral galaxies, \citet{Ferr02} measured a relationship between $v_{\rm c}$ and $\sigma_{\rm c}$.  Combining these measurements with the compiled relationship between SMBH mass and $\sigma_{\rm c}$ presented by \citet{Ferr02se} reveals that observations are consistent with a redshift-independent $M_{\rm BH} \propto v_{\rm c}^5$ relation.

\citet{WL03} proposed a mechanism for black-hole-mass regulation that would result in a $M_{\rm BH} \propto v_{\rm c}^5$ relation between central-black-hole mass and disc circular velocity for all redshifts.  They postulated that a black hole ceases to accrete when the power radiated by the accretion exceeds the binding energy of the host galactic disc divided by the dynamical time of the disc.  Assuming that the accretion disc shines at its Eddington luminosity, the black hole stops growing when
\begin{equation}
M_{\rm BH} = 1.9 \times 10^8 \left(\frac{F_{\rm q}}{0.07}\right)\left(\frac{v_{\rm c}}{350\mbox{ km s$^{-1}$}}\right)^5 \sm ,
\label{wlmbh}
\end{equation}   
where $F_q$ is the fraction of the radiated power which is transferred to gas in the disc.  Setting $F_q$ to 0.07 brings equation (\ref{wlmbh}) into agreement with the observations presented by \citet{Ferr02}.    

The final step in the determination of a halo--black-hole-mass relation is to connect the circular velocity to the halo mass via the virial velocity \citep{BL01}.  The simplest possible assumption is that the circular velocity of the disc equals the virial velocity of the halo.  This assumption is made by \citet{WL03}, and we assume that $v_{\rm c} = v_{\rm vir}$ throughout this paper.  However, different relations between $v_{\rm c}$ and $v_{{\rm vir}}$ have been proposed and can significantly impact the final $M_{\rm BH}$-$M_{\rm halo}$ relation \citep[see][]{Ferr02}.  

\begin{figure}
\begin{center}
\resizebox{8cm}{!}{\includegraphics{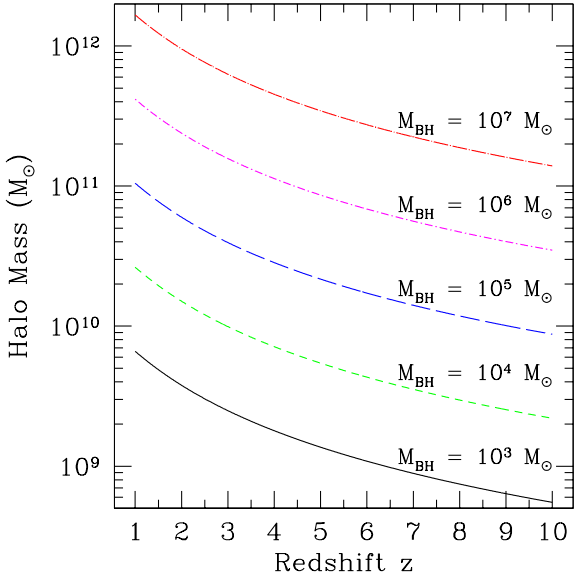}}
\caption{The masses of haloes that contain supermassive black holes of mass $10^3, 10^4, 10^5, 10^6 \mbox{ and }10^7$ $\sm $, according to the $M_{\rm BH}$--$M_{\rm halo}$ relation proposed by \citet{WL03} for a flat \lcdm universe with $\Om =0.27$.  This relation is normalized to fit local observations and assumes that the disc circular velocity equals the virial velocity.}  
\label{fig:BHmass}
\end{center}
\end{figure}

Assuming that $v_{\rm c} = v_{\rm vir}$, the halo mass then becomes a redshift-dependent function of the mass of the central black hole:
\begin{equation}
\frac{M_{\rm halo}}{10^{12} \sm } = 10.5\left(\frac{\Om^0}{\Om(z)}\frac{\Delta_{\rm c}}{18\pi^2}\right)^{-\frac{1}{2}}(1+z)^{-\frac{3}{2}}\left(\frac{M_{\rm BH}}{10^{8} \sm }\right)^{\frac{3}{5}},
\label{MbhMhalo}
\end{equation}
where $\Om(z)$ the matter density in units of the critical density as a function of redshift, $\Om^0 \equiv \Om(z=0)$, and $\Delta_{\rm c}$ is the nonlinear over-density at virialization for a spherical top-hat perturbation in a \lcdm universe:
\begin{equation}
\Delta_{\rm c} = 18\pi^2 + 82[\Om(z)-1]-39[\Om(z)-1]^2.
\label{Delc}
\end{equation}
Figure \ref{fig:BHmass} shows the masses of haloes that contain supermassive black holes of several masses.  Citing the fact that the largest haloes observed at low redshifts appear to contain galaxy clusters with no central black holes, \citet{WL03} argue that supermassive-black-hole growth was complete by $z\sim1$ and that local SMBH masses reflect the limiting values at that redshift.  Consequently, when determining the mass of a halo that contains a black hole of a given mass, we use the $z=1$ value of equation (\ref{MbhMhalo}) for all redshifts less than one.        

Some calculations of the LISA SMBH-merger event rate impose a minimum halo virial temperature instead of a minimum black-hole mass when calculating the lower mass bound on haloes that contribute to the SMBH merger rate \citep{WL03gw, RW05}.  This constraint reflects the fact that supermassive black holes only form when the gas within dark-matter haloes can cool.  However, the relation between virial temperature and virial mass \citep{BL01} may be be used to eliminate the halo mass in equation (\ref{MbhMhalo}) in favour of the virial temperature. The redshift-dependent terms cancel, leaving a redshift-independent relation between black-hole mass and halo virial temperature:
\begin{equation} 
M_{\rm BH} = (267 \mbox{ }\sm )\mbox{ }h^{-5/3} \left(\frac{T_{\rm vir}}{1.98\times 10^4 \mbox{ K}}\right)^{5/2}.
\label{MbhTvir}
\end{equation}
Therefore, defining the minimum halo mass by a minimum halo virial temperature is nearly equivalent to defining it by a minimum black-hole mass via equation (\ref{MbhMhalo}).  For example, requiring that the halo's virial temperature be significantly higher than the temperature of the intergalactic medium, $T_{\rm vir}\gsim 10^5 K$ \citep{WL03gw}, corresponds to imposing a minimum black-hole mass of $2.6\times 10^4 \sm $.  The only discrepancy occurs when $z<1$, because we assume that the $M_{\rm BH}$--$M_{\rm halo}$ relation is fixed for redshifts less than one, while $T_{\rm vir}$ is still redshift dependent.  However, we shall see that nearly all SMBH mergers occur at redshifts greater than one, so this difference is negligible.     

\section{EPS merger theory and LISA event rates}
\label{sec:EPS}
\subsection{Review of EPS merger theory} 
\label{sec:EPSreview}
The first pillar of extended Press--Schechter (EPS) merger theory is the Press--Schechter halo mass function \citep{PS74}, which gives the number of haloes with masses between $M$ and $M+{\rm d}M$ per comoving volume:
\begin{equation}
\frac{{\rm d}n_{\mathrm{halo}}}{{\rm d}\ln M} = \sqrt{\frac{2}{\pi}}\frac{\rho_0}{M} \left(\left|\frac{{\rm d} \ln \sigma}{{\rm d} \ln M}\right|_{M}\right) \frac{\delta_{\mathrm{coll}}}{\sigma(M,z)}
\exp\left[\frac{-\delta_{\mathrm{coll}}^2}{2\sigma^2(M,z)}\right],
\label{PSn}
\end{equation}
 where $\rho_0$ is the background matter density today, $\delta_{\mathrm{coll}}$ is the critical linear over-density for collapse in the spherical-collapse model, and $\sigma(M, z)$ is the root variance of the linear density field at redshift $z$ in spheres containing mass $M$ on average.  In a $\Lambda$CDM universe, $\delta_{\mathrm{coll}}$ deviates slightly from its Einstein-de Sitter value of $\sim 1.686$ when the cosmological constant begins to dominate the energy density of the Universe \citep{KS96,WK03}.  In this work, the fitting function obtained by \citet{KS96} was used to approximate $\delta_{\mathrm{coll}}$.  When calculating $\sigma(M, z)$, we assumed a scale-invariant primordial power spectrum and we used the transfer function provided by \cite{EH98}.

The second pillar of EPS merger theory is the merger probability function derived by \citet{LC93}, which gives the probability that a halo of mass $M_1$ will become a halo of mass $M_{\rm f} \equiv M_1+M_2$ per unit time, per unit acquired mass: 
\begin{eqnarray}
\frac{{\rm d}^2 p}{{\rm d}t\mbox{ }{\rm d}M_2} &=& \frac{1}{M_{\rm f}}\sqrt{\frac{2}{\pi}}\left|\frac{\dot{\delta}_{\mathrm{coll}}}{\delta_{\mathrm{coll}}}-\frac{\dot{D}(t)}{D(t)}\right|\left(\left|\frac{{\rm d} \ln \sigma}{{\rm d} \ln M}\right|_{M_{\rm f}}\right) \nonumber\\
&&\times\frac{\delta_{\mathrm{coll}}}{\sigma(M_{\rm f},z)}\left(1-\frac{\sigma^2(M_{\rm f},z)}{\sigma^2(M_1,z)}\right)^{-3/2}\nonumber\\
&&\times\exp\left[{\frac{-\delta_{\mathrm{coll}}^2}{2}\left(\frac{1}{\sigma^2(M_{\rm f}, z)} - \frac{1}{\sigma^2(M_1, z)}\right)}\right].
\label{dpdMdt}
\end{eqnarray}
In this expression, $D(t)$ is the linear growth function, and a dot denotes differentiation with respect to time.  

Equation (\ref{dpdMdt}) is usually interpreted as the differential probability that a given halo of mass $M_1$ will merge with a halo of mass between $M_2$ and $M_2+{\rm d}M_2$ per unit time, per increment mass change.  Thus equation (\ref{dpdMdt}) already includes information about the abundance of haloes of mass $M_2$, but not the abundance of haloes of mass $M_1$.  Following \citetalias{BKH05}, it is revealing to examine a different quantity, which does not differentiate between the two merging haloes: the rate of mergers between haloes of masses $M_1$ and $M_2$ per comoving volume,
\begin{eqnarray}
R(M_1,M_2,t) &\equiv& \frac{\mbox{Number of $M_1 + M_2$ Mergers}}{{\rm d}t\mbox{  } {\rm d}(\mbox{Comoving Volume})},\nonumber\\
&=& \left(\frac{{\rm d}n(M_1;t)}{{\rm d}M_1}\right)\left(\frac{{\rm d}^2 p}{{\rm d}t \mbox{  }{\rm d}M_2}\right)\mbox{ }{\rm d}M_1 \mbox{  }{\rm d}M_2.
\label{Rdef}
\end{eqnarray} 
The EPS self-inconsistency documented by \citetalias{BKH05} manifests itself here.  Although $R(M_1,M_2,t)$ must be symmetric in its mass arguments by definition, equation (\ref{Rdef}) is not symmetric under exchange of $M_1$ and $M_2$.

The mass asymmetry of EPS merger theory becomes most transparent when one defines a new function: the merger kernel.  From its definition, it is apparent that  $R(M_1,M_2,t)$ should be proportional to the number densities of both haloes involved in the merger.  Extracting this dependence defines the merger kernel $Q(M_1,M_2,t)$:
\begin{eqnarray}
R(M_1,M_2,t) &\equiv& \left(\frac{{\rm d}n(M_1;t)}{{\rm d}M_1}\right)\left(\frac{{\rm d}n(M_2;t)}{{\rm d}M_2}\right)\nonumber\\
&&\times Q(M_1,M_2, t)\mbox{ }{\rm d}M_1\mbox{ }{\rm d}M_2.
\label{Qdef}
\end{eqnarray}
In addition to isolating the source of the mass-asymmetry in EPS merger theory, the merger kernel enters into the coagulation equation which is inverted to obtain BKH merger rates, as described in Section \ref{sec:BKH}.

\begin{figure}
\begin{center}
\resizebox{8cm}{!}{\includegraphics{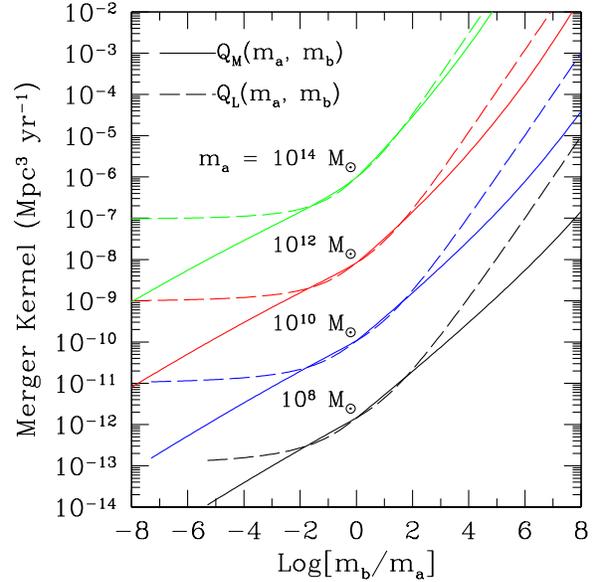}}
\caption{The two EPS merger kernels for $z=0$. Here, $Q_{\rm M}$ is the Lacey--Cole merger kernel with the more massive halo as the first argument, and $Q_{\rm L}$ is the same kernel with the less massive halo as the first argument.  Results are shown for a flat \lcdm universe with $\Om = 0.27$, $h=0.72$ and $\sigma_8 = 0.9$.}
\label{fig:QMvsQL}
\end{center}
\end{figure}

The EPS merger kernel $Q(M_1,M_2)$ is the probability function given by equation (\ref{dpdMdt}) divided by the number density of haloes of mass $M_2$ given by equation (\ref{PSn}).  In effect, EPS merger theory includes two distinct merger kernels, depending on the order of the mass arguments.  Thus, we define two mass-symmetric merger kernels: $Q_{\rm M}(M_1,M_2)$ equals the EPS merger kernel with the more massive halo as the first argument, while $Q_{\rm L}(M_1,M_2)$ equals the EPS merger kernel with the less massive halo as the first argument.  Figure \ref{fig:QMvsQL} illustrates the differences in the merger kernels $Q_{\rm M}$ and $Q_{\rm L}$.  Note that neither $Q_{\rm M}(M_1, M_2)$ nor $Q_{\rm L}(M_1, M_2)$ are viable candidates for the true halo merger kernel because they are not smooth functions of halo mass.  They are useful because they expose the ambiguities hidden in applications of EPS merger theory.

In order to avoid double counting mergers when calculating a merger rate, it is common to restrict one mass argument to be larger than the other.  Using the standard expression for the Lacey--Cole merger probability function, as given by equation (\ref{dpdMdt}), in such calculations is equivalent to using $Q_{\rm M}(M_1, M_2)$ or $Q_{\rm L}(M_1, M_2)$.  Specifically, \citet{Haeh94} effectively used $Q_{\rm L}$ to predict an event rate for LISA, while \citet{WL03gw, RW05} effectively used $Q_{\rm M}$.  Using the other version of the EPS merger kernel in either of these calculations would have yielded different results, as we show in Section \ref{sec:EPSResults}.  More generally, any application of the Lacey--Cole merger probability function uses some mixture of $Q_{\rm M}$ and $Q_{\rm L}$, and changing the mixture will change the result of the calculation.   
 
\subsection{LISA event rates from EPS theory} 
\label{sec:EPSResults}   

The rate of SMBH mergers per comoving volume follows from the rate of halo mergers per comoving volume given in equation (\ref{Rdef}):
\begin{eqnarray}
{\cal{N}}(z) \equiv \frac{1}{2}\int_{M_{\mathrm{min}}}^{\infty}\int_{M_\mathrm{min}}^{\infty}& \left(\frac{{\rm d}n(M_1,z)}{{\rm d}M}\right)\left(\frac{{\rm d}n(M_2,z)}{{\rm d}M}\right)\nonumber\\&\times Q(M_1,M_2,z)\mbox{ }{\rm d}M_1\mbox{ }{\rm d}M_2,
\label{scriN}
\end{eqnarray}
where $M_{\rm min}$ is the minimum halo mass that contains a black hole massive enough to be detected when it mergers with a black hole of equal or greater mass.  The factor of 1/2 accounts for the double counting of mergers.  Some calculations \citep[e.g.][]{RW05} only include mergers between haloes with mass ratios less than three and so integrate $M_2$ from $M_1/3$ to $3M_1$.  This restriction is motivated by dynamical-friction calculations that indicate that when a halo merges with a halo less than a third of its size, it takes longer than a Hubble time for their central black holes to merge \citep*{CMG99}.  However, recent numerical simulations indicate that this restriction may be too strict; when gas dynamics are included, SMBHs with host-galaxy-mass ratios greater than three merge within a Hubble time \citep{Kaz05}.  We do not impose any restrictions on the halo-mass ratios, so our event rates are upper bounds arising from the assumption that every halo merger in which both haloes contain a SMBH results in a SMBH merger.

\begin{figure}
\begin{center}
\resizebox{8cm}{!}{\includegraphics{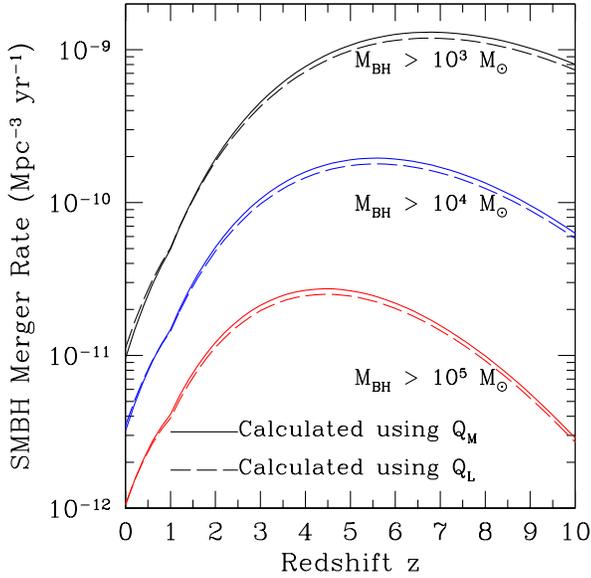}}
\caption{The rate of SMBH mergers per comoving volume where both merging black holes have a mass greater than $10^3 \sm $, $10^4 \sm $ and $10^5 \sm $.   The solid (dashed) lines show the results when the first argument of the Lacey--Cole merger kernel is the more (less) massive halo.  Results are shown for a flat \lcdm universe with $\Om = 0.27$, $h=0.72$, and $\sigma_8 = 0.9$.}
\label{fig:scriNMbh}
\end{center}
\end{figure}

Since LISA should observe mergers between two SMBHs with masses greater than $10^4 \sm $ out to redshifts of at least eight, we generally use this minimum black hole mass to determine $M_{\rm min}$.  The corresponding rates of SMBH mergers per comoving volume are shown in Figure \ref{fig:scriNMbh}, as well as the rates which correspond to different choices for the minimum mass of a SMBH.  Both versions of $\scriN$ are shown to illustrate the difference between the two Lacey--Cole merger kernels.  The crimp in $\scriN(z)$ at $z = 1$ reflects the transition from a constant $M_{\rm min}$ (evaluated at $z=1$) to the redshift-dependent form given by equation (\ref{MbhMhalo}). 

\begin{figure}
\begin{center}
\resizebox{8cm}{!}{\includegraphics{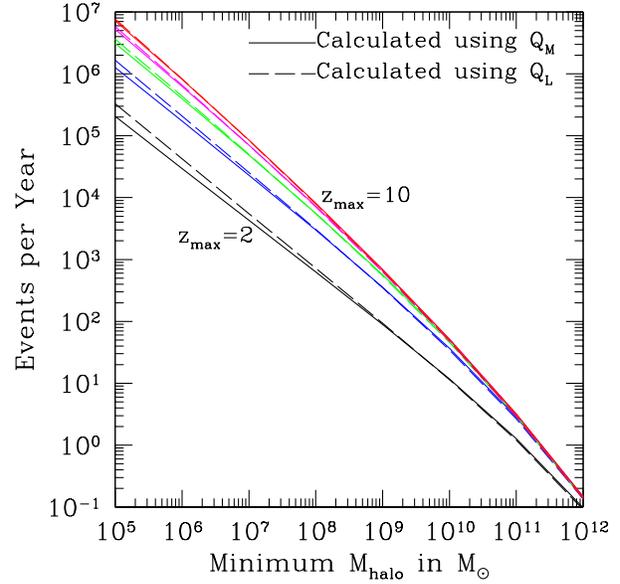}}
\caption{The gravitational-wave event rate from SMBH mergers as a function of the minimum halo mass that contains a SMBH large enough to produce a detectable signal when it merges.  Mergers at redshifts up to $z_{\rm max}$ were included in this rate, and the five pairs of lines correspond to $z_{\rm max}=2,4,6,8,10$.  The solid (dashed) lines show the results when the first argument of the Lacey--Cole merger kernel is the more (less) massive halo.  Results are shown for a flat \lcdm universe with $\Om = 0.27$, $h=0.72$, and $\sigma_8 = 0.9$.}
\label{fig:GWEMmin}
\end{center}
\end{figure}

\begin{figure}
\begin{center}
\resizebox{8cm}{!}{\includegraphics{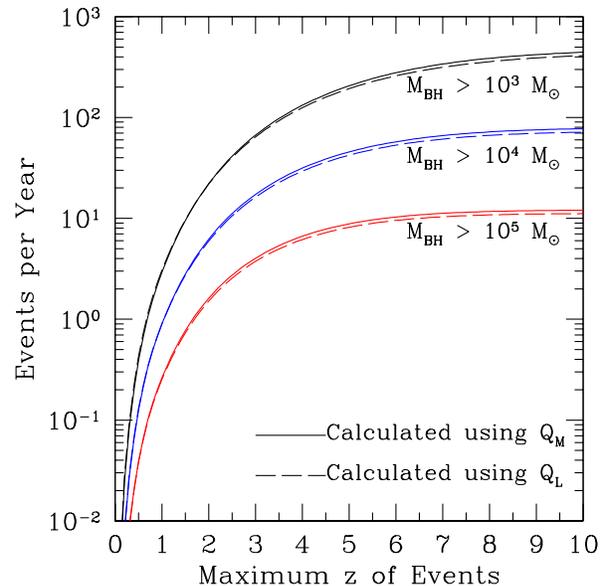}}
\caption{The gravitational-wave event rate from SMBH mergers as a function of the maximum redshift of a detectable merger.  Only mergers in which both black holes have masses greater than the given lower bound are included.  The solid (dashed) lines show the results when the first argument of the Lacey--Cole merger kernel is the more (less) massive halo.  Results are shown for a flat \lcdm universe with $\Om = 0.27$, $h=0.72$, and $\sigma_8 = 0.9$.}
\label{fig:GWEvsZ}
\end{center}
\end{figure}

Once the rate $\scriN(z)$ of SMBH mergers per volume is known, equation (\ref{dBdz}) may be integrated over redshift to obtain an event rate for LISA.  Figure \ref{fig:GWEMmin} shows the LISA event rate as a function of the minimum halo mass that contains a black hole large enough to emit an observable signal.  Figure \ref{fig:GWEvsZ} shows the event rate as a function of the maximum redshift of a detectable merger.  In Figure \ref{fig:GWEvsZ}, $M_{\rm min}$ is the mass of a halo that contains a black hole more massive than $10^3$, $10^4$, or $10^5 \sm $ as determined by the $M_{\rm BH}-M_{\rm halo}$ relation given by equation (\ref{MbhMhalo}).  These rates correspond to the $\scriN$ results depicted in Figure \ref{fig:scriNMbh}.  Examination of these results reveals that increasing $z_{\rm max}$ beyond $z_{\rm max}=6$ has little effect on the event rate when $M_{\rm min}$ is greater than $10^{9} \sm $, as is the case when equation (\ref{MbhMhalo}) is used to obtain the value of $M_{\rm min}$ which corresponds to a minimum black-hole mass of $10^4 \sm $.  The levelling of the event rate for $z_{\rm max}\gsim6$ indicates that SMBH mergers are very rare at higher redshifts and that the event rate is dominated by mergers that occur at redshifts $z\lsim 6$.  Therefore, the upper bound on LISA's sensitivity to larger SMBH mergers at high redshifts will have little effect on the event rate.  

The event rates shown in Figures \ref{fig:GWEMmin} and \ref{fig:GWEvsZ} differ significantly from those calculated by \citet{WL03gw} and \citet{RW05}.  Our event rates are generally much higher than the event rates reported by \citet{WL03gw} because we do not exclude mergers between haloes with mass ratios greater than three from our SMBH merger rate.  The event rates calculated by \citet{RW05} are even lower because they do not assume that all haloes contain galaxies.  The one case where our event rates are not substantially higher than those derived by \citet{RW05} is when the minimum black-hole mass is taken to be very high ($M_{\rm BH} \gsim 10^5 \sm$).  In that case, the minimum halo mass is so high that nearly all mergers involve haloes of similar masses ($M_{\rm halo}\sim 10^{11} \sm$), and the galaxy-occupation fraction derived by \citet{RW05} indicates that nearly all haloes of this size contain galaxies for redshifts greater than three, so our event rate of 12 per year is very similar to the result of the more sophisticated treatment of \citet{RW05}. \footnote{When we attempted to reproduce the differential event rates calculated by \citet{Haeh94}, we found that our rates are roughly a factor of two lower.  After extensive review and two independent calculations, we were unable to find any errors in our analysis.} 

Event rates obtained from both versions of the EPS merger kernel are shown in Figures \ref{fig:GWEMmin} and \ref{fig:GWEvsZ}.  The differences between these results reveal the type of mergers that dominate the calculation.  As shown in Figure \ref{fig:GWEvsZ}, the event rate obtained from $Q_{\rm M}$ is slightly higher than the rate obtained from $Q_{\rm L}$, indicating that mergers with halo mass ratios less than $10^2$ are dominating the sum (see Figure \ref{fig:QMvsQL}).  For a constant value of $M_{\rm min}=10^5 \sm$, the difference between the two versions decreases as the maximum redshift increases, as shown in Figure \ref{fig:GWEMmin}.  This convergence indicates that the contribution from mergers between haloes of greatly unequal masses to the event rate dwindles as redshift increases.  Since the lower bound on halo mass is constant with redshift, a decrease in unequal-mass mergers reflects a decrease in the population of larger haloes.  The Press-Schechter mass function implies that the largest halo that is common at a given redshift is given by the function $M_*(z)$, which is defined by the relation $\sigma(M_*,z) \equiv \delcoll(z)$.  When $M>M_*(z)$, the exponential term in equation (\ref{PSn}) dominates, and the number density of such halos is exponentially suppressed. $M_*(z)$ decreases with redshift, reflecting the fact that at early times, massive halos had yet to form.  Due to the exponential decline in the number density of haloes greater than $M_*(z)$, there is an effective upper bound to the integrals in equation (\ref{scriN}), which defines $\scriN(z)$.  This upper bound on halo mass follows $M_*$ and is less than 100 times greater than $M_{\min}$ at the redshifts which dominate the merger rate.    

\begin{figure}
\begin{center}
\resizebox{8cm}{!}{\includegraphics{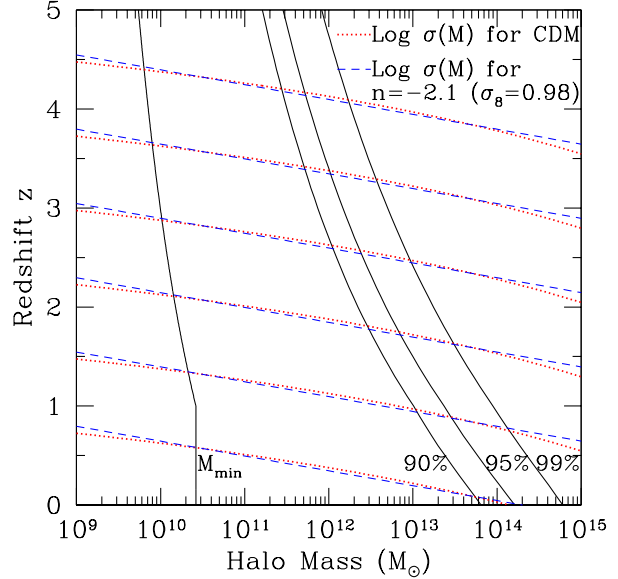}}
\caption{The halo mass range that dominates the rate of SMBH mergers per comoving volume.  The three curves marked with percentages define the upper bounds of mass ranges that account for 90\%, 95\% and 99\% of $\scriN$.  Here, $M_{\rm min}$ is the mass of a halo that contains a SMBH of mass $10^4\sm $.  The dotted curves are plots of $\log\sigma(M)$ with arbitrary normalizations. Results are shown for a flat \lcdm universe with $\Om = 0.27$ and $\sigma_8 = 0.9$. The dashed lines are plots of $\log\sigma(M)$ for a power-law power spectrum with $n=-2.1$ and $\sigma_8=0.9843$, which is the best linear fit to $\log\sigma$ over the mass range between $M_{\rm min}$ and the 99\% curve for $z\leq5$. }
\label{fig:RangeWithSig}
\end{center}
\end{figure}

 The relevant mass range may be quantified by considering the ratio, 
\begin{eqnarray*}
C(U) \equiv  \frac{ \frac{1}{2}\int_{M_{\mathrm{min}}}^{U} {\rm d}M_1 \int_{M_\mathrm{min}}^{U} {\rm d}M_2 \left(\frac{{\rm d}n}{{\rm d}M_1}\right)\left(\frac{{\rm d}n}{{\rm d}M_2}\right)Q(M_1,M_2)}{ \frac{1}{2}\int_{M_{\mathrm{min}}}^{\infty} {\rm d}M_1 \int_{M_\mathrm{min}}^{\infty} {\rm d}M_2 \left(\frac{{\rm d}n}{{\rm d}M_1}\right)\left(\frac{{\rm d}n}{{\rm d}M_2}\right)Q(M_1,M_2)},
\end{eqnarray*}
where the $z$-dependence of all quantities has been suppressed.  Using the standard Lacey--Cole merger kernel when evaluating $C(U)$ is equivalent to using the arithmetic mean of $Q_{\rm M}$ and $Q_{\rm L}$.  Figure \ref{fig:RangeWithSig} shows the values of $U$ for $C=0.9,0.95$ and $0.99$.  Since the dominant halo mass range is so narrow, it is possible to find a power-law power spectrum that accurately approximates the value of $\sigma(M)$ over this mass range, and both the exact and the approximate $\sigma(M)$ are shown in Figure \ref{fig:RangeWithSig}.  This approximation will allow us to apply BKH merger theory to the calculation of LISA event rates in Section \ref{sec:BKHvsEPS}.

\section{BKH merger theory}
\label{sec:BKH}

\subsection{Solving the coagulation equation}
A merger kernel that preserves the Press--Schechter (PS) halo mass distribution must satisfy the Smoluchowkski coagulation equation \citep{Smol16}, which simply states that the rate of change in the number of haloes of mass $M$ equals the rate of creation of such haloes through mergers of smaller haloes minus the rate haloes of mass $M$ merge with other haloes.  Adopting the shorthand $n(M)$ for the PS halo number density per interval mass and suppressing the redshift dependence of all terms, the coagulation equation is
\begin{eqnarray}
\frac{{\rm d}}{{\rm d}t}n(M)&=&\frac{1}{2}\int^M_0 n(M^\prime)n(M-M^\prime)Q(M^\prime,M-M^\prime)\mbox{ }{\rm d}M^\prime \nonumber\\
&&-n(M)\int^\infty_0 n(M^\prime)Q(M,M^\prime)\mbox{ }{\rm d}M^\prime, 
\label{coag}
\end{eqnarray}
where $Q(M_1,M_2,z)$ is the desired merger kernel.  The first term on the right-hand side is the rate of mergers per comoving volume that create a halo of mass $M$.  The second term is the rate of mergers involving a halo of mass $M$ per comoving volume --- these mergers effectively destroy haloes of mass $M$.  

\citetalias{BKH05} numerically invert the coagulation equation for $Q$ for power-law density power spectra $P(k) \propto k^n$.  When the density power spectrum is a power law, $\sigma(M)$ is proportional to $M^{-(3+n)/6}$.  Since the redshift-dependence of the PS mass function enters via the ratio $\delcoll(z)/\sigma(M,z) = (M/M_*)^{(3+n)/6}$, the $z$-dependence of the PS mass function may be eliminated by expressing the masses in units of $M_*(z)$.  For a judicious choice of time variables, differentiating the PS mass function introduces no $z$-dependence, and the coagulation equation becomes redshift-invariant.  Consequently, the coagulation equation only has to be inverted once, for the resulting merger kernel $Q(M_1/M_*,M_2/M_*)$ is applicable to all redshifts.  This simplification is only possible when the power spectrum is a power law.  For more complicated spectra, the coagulation equation will have to be solved at multiple redshifts.

When they numerically solve the coagulation equation on a discrete grid, \citetalias{BKH05} require that the merger kernel be symmetric in its two mass arguments.  However, this restriction is not sufficient to determine $Q$ uniquely from the coagulation equation.  On an $N \times N$ mass grid, the coagulation equation becomes $N$ equations for the $N$ possible values of $M$.  Meanwhile, the symmetric $Q$ matrix on the grid, $Q_{ij} = Q(M_i, M_j)$, has $N(N+1)/2$ independent components. To break the degeneracy, \citetalias{BKH05} impose a regularization condition.  By minimizing the second derivatives of $Q$, they find the smoothest, non-negative kernel that solves the coagulation equation.  

\subsection{BKH merger rates for power-law power spectra}  

In Section \ref{sec:EPSResults}, we demonstrated that the rate of SMBH mergers per comoving volume is dominated by mergers between haloes in a very limited mass range, as shown in Figure \ref{fig:RangeWithSig}.  The $\sigma(M)$ curves in Figure \ref{fig:RangeWithSig} show that it is possible to accurately approximate $\sigma(M)$ over the relevant mass range as originating from a power-law power spectrum.  We consider a power-law fit for $\sigma(M)$ that extends over all masses that fall within the 99\% mass range at any redshift less than five.  The fit has a lower mass bound of $5.44\times10^9 \sm $, which is the value of $M_{\rm min}$ at $z=5$, and extends to a mass of $4.26\times10^{14} \sm $.  Over this range, $\sigma(M)$ is best fit by spectral index $n = -2.1$ normalized so that $\sigma_8 = 0.9843$, as shown by the dashed lines in Figure \ref{fig:RangeWithSig}.  This $n = -2.1$ power-law approximation of $\sigma(M)$ is accurate to within 16\% over this mass range.  We chose to fit the mass range for $z\lsim5$ because the SMBH merger rate peaks at redshifts less than five when the minimum black-hole mass is greater than $10^4 \sm $, as shown in Figure \ref{fig:scriNMbh}.  Also, when the mass range is lowered, the best-fitting spectral index decreases, and BKH merger rates have not been obtained for $n<-2.2$.     
 
The density power spectrum enters the EPS merger kernel only through $\sigma(M)$, so any power-law approximation that accurately models $\sigma(M)$ for $M_1$, $M_2$, and $M_{\rm f} = M_1+M_2$ will accurately model the Lacey--Cole merger kernel $Q(M_1,M_2,z)$.  Unfortunately, the same is not necessarily true for the BKH merger kernels obtained by inverting the coagulation equation.  Since the coagulation equation [equation (\ref{coag})] involves integrals over all masses and is solved for all masses on the grid, the solution $Q(M_1,M_2,z)$ is dependent on $\sigma(M)$ over all masses and not just the arguments of the kernel.  Therefore, while the power-law approximation accurately reflects the full \lcdm result for EPS merger theory, the BKH merger rates obtained for the same power law may differ greatly from the merger rates that solve the coagulation equation for a \lcdm universe.  However, since the coagulation equation has not been solved for a \lcdm power spectrum, we compare the EPS merger rates to the BKH merger rates for the same power law.  This comparison demonstrates how the BKH merger rates differ from the EPS rates, but should not be considered a definitive description of merger rates in a \lcdm universe.  

BKH merger kernels for a power-law power spectrum with $n = -2.1$ were obtained by inverting the coagulation equation on a $91\times91$ grid of logarithmically-spaced $M/M_*$ values ranging from $10^{-12}$ to $3000$.  For $M/M_*$ values greater than $10^{-8}$, the merger kernel values are not dependent on grid resolution, which indicates that the kernel is a numerically robust solution of the discretized coagulation equation for masses above $10^{-8}M_*$.  The $M_{\mathrm{BH}}$--$M_{\mathrm{halo}}$ relation [equation (\ref{MbhMhalo})] implies that SMBHs with masses greater than $10^3 \sm $ and redshifts less than ten reside in haloes with masses greater than $10^8 \sm$, while the $z=0$ value of $M_*$ for the $n = -2.1$ power-law power spectrum is $6 \times 10^{12} \sm$.  Therefore, for all haloes that contain SMBHs capable of producing a gravitational wave signal detectable by LISA, $M/M_* \gsim 10^{-5}$, so the lower mass bound on reliable kernel values is of no concern.  

Unfortunately, the same is not true for the upper bound on $M/M_*$.  The upper bound on the halo masses which contribute to the SMBH merger rate $\cal{N}$ in EPS theory, shown in Figure \ref{fig:RangeWithSig}, extends to $M/M_* \gsim 10^5$ for $z \gsim 5$.  However, extending the mass grid to higher values of $M/M_*$ introduces numerical noise that prevents the kernels from converging as grid resolution is increased.  Therefore, we must extrapolate the BKH kernel to higher masses.  We bilinearly extrapolate the logarithm of the kernel with respect to the logarithms of its mass arguments.  When used to extrapolate from a grid with $M/M_* < 100$, this recovers the kernel to within a factor of two.  Moreover, ignoring mergers of haloes with $M/M_* > 3000$ only slightly affects the gravitational-wave event rate calculated from the BKH merger rates: the event-rate reduction is less than $3\%$.  Therefore, the errors introduced by our extrapolation of the BKH merger kernel are negligible.       

\begin{figure}
\begin{center}
\resizebox{8cm}{!}{\includegraphics{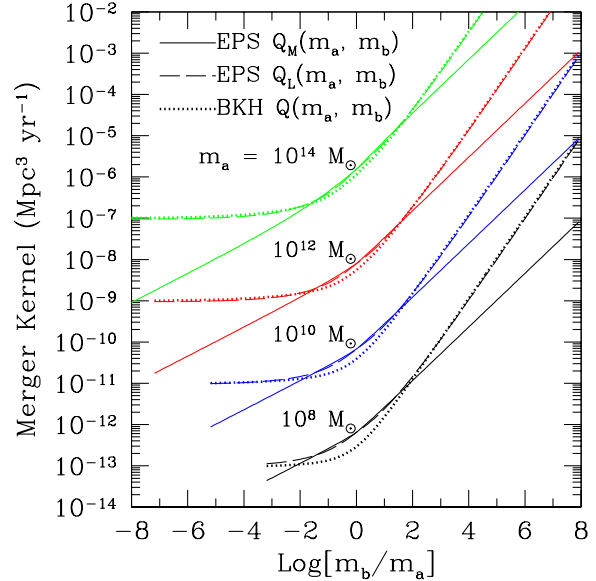}}
\caption{The two EPS merger kernels and the BKH merger kernel for a $n=-2.1$ power-law power spectrum at $z=0$. Here, $Q_{\rm M}$ is the Lacey--Cole merger kernel with the more massive halo as the first argument, and $Q_{\rm L}$ is the same kernel with the less massive halo as the first argument.  Results are shown for $\Om = 0.27$, $h=0.72$ and $\sigma_8 = 0.98$.  The low-mass cut-off of the curves arises from the $M/M_*\gsim 10^{-8}$ bound on the BKH merger kernel.}
\label{fig:PlQmVsQl}
\end{center}
\end{figure}

The differences between the BKH merger kernel and both versions of the EPS merger kernel are illustrated by Figure \ref{fig:PlQmVsQl}.  The BKH merger  kernel is less than both EPS kernels when the masses of the merging haloes are similar, and the difference increases as the haloes get smaller.  For mergers between haloes with mass ratios greater than $10^2$, the BKH merger kernel is nearly equal to the EPS kernel with the least-massive halo as the first argument ($Q_{\rm L}$) for all masses.  Therefore, for an $n=-2.1$ power-law power-spectrum, $Q_{\rm L}$ comes closer to solving the coagulation equation than $Q_{\rm M}$.  

\section{Comparison of LISA event rates from BKH and EPS merger theories}
\label{sec:BKHvsEPS}
\begin{figure}
\begin{center}
\resizebox{8cm}{!}{\includegraphics{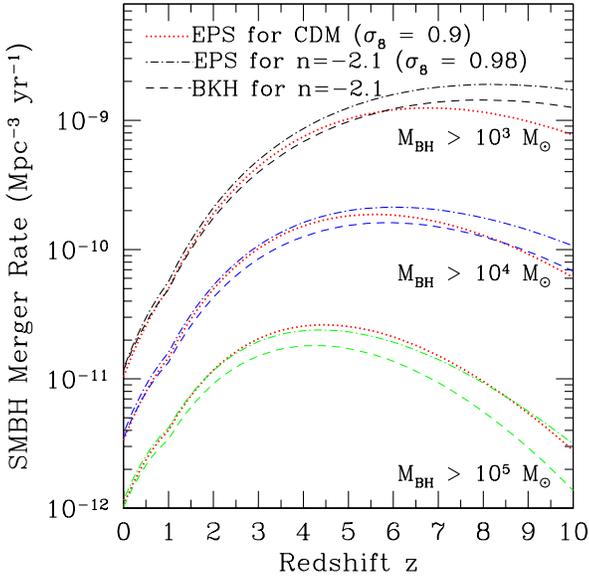}}
\caption{The rate of SMBH mergers per comoving volume where both merging black holes have a mass greater than $10^3 \sm $, $10^4 \sm $, or $10^5 \sm $.  The dotted line shows the EPS merger kernel for a \lcdm power spectrum with $\sigma_8=0.9$.  The dot-dashed curves are the results derived from EPS theory for a power-law approximation with $n=-2.1$ and $\sigma_8=0.98$.  The dashed curves are the BKH results for the same power law and normalization.  These results all assume a flat \lcdm universe with $\Om = 0.27$ and $h=0.72$.}
\label{fig:PLScriN}
\end{center}
\end{figure}

Since the BKH merger kernels for haloes of nearly equal masses are smaller than the EPS kernels for the same spectral index, applying EPS merger theory may over-estimate the LISA event rate.  Figure \ref{fig:PLScriN} shows the rate $\scriN$ of SMBH mergers per comoving volume for the power-law model discussed in the previous Section.  For comparison, the EPS results for a \lcdm universe are also shown as dotted curves (these are the arithmetic means of the corresponding solid and dashed curves in Figure \ref{fig:scriNMbh}).  However, it is important to remember that although the power-law models may accurately approximate the \lcdm results in the EPS theory, the same should not be assumed for the BKH merger rates.  The BKH merger rates should only be compared to the EPS rates for the same power law.  

The discrepancy between the power-law EPS results and the \lcdm curves at high redshifts is attributable to the power-law halo number density, which is much greater than the \lcdm halo number density for masses below $10^{11} \sm $ at these redshifts.  The same mass function is used to calculate the merger rate in BKH theory, so when the power-law merger rate is higher than the \lcdm rate in EPS theory, it is reasonable to assume that the same is true for the rate derived from BKH theory.  Figure \ref{fig:PLScriN} also shows that the predictions for the SMBH merger rate from the BKH and EPS merger theories diverge with increasing redshift.  In Section \ref{sec:EPSResults}, we showed that as redshift increases, nearly equal-mass halo mergers dominate the event rate.  The differences between the BKH merger kernel and the EPS kernel are greatest when the masses of the merging haloes are nearly equal, so as these mergers dominate the event rate at high redshifts, the BKH and EPS merger rates diverge.       

\begin{figure}
\begin{center}
\resizebox{8cm}{!}{\includegraphics{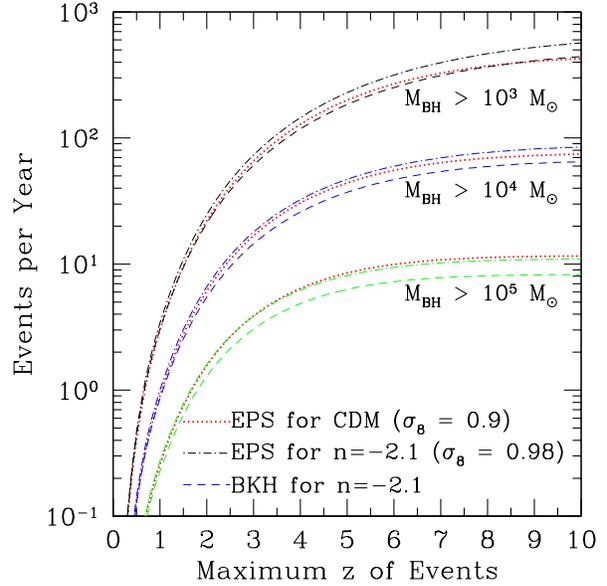}}
\caption{The gravitational-wave event rate from SMBH mergers as a function of the maximum redshift of a detectable merger.  Only mergers in which both black holes have a mass greater than $10^3 \sm $, $10^4 \sm $, or $10^5 \sm $ are included.  The dotted line shows the EPS merger kernel for a \lcdm power spectrum with $\sigma_8=0.9$.  The dot-dashed curves are the results derived from EPS theory for a power-law approximation: $n=-2.1$ and $\sigma_8 = 0.98$.  The dashed curves are the BKH results for the same power law and normalization.  These results all assume a flat \lcdm universe with $\Om = 0.27$ and $h=0.72$.}
\label{fig:PLGWE}
\end{center}
\end{figure}

Figure \ref{fig:PLGWE} illustrates the potential consequences BKH merger theory has for the SMBH merger event rate observed by LISA.  The difference between the BKH and EPS merger kernels for the same spectral index leads to a fairly substantial difference in the resulting event rates for LISA.  For realistic values of the maximum redshift of a detectable merger ($z_{\rm max} \gsim 5$), the EPS prediction is about thirty percent higher than the BKH prediction for the $n=-2.1$ power-law approximation.  If the BKH merger kernel for a full \lcdm power spectrum preserves the ratio of the BKS and EPS event rates for this spectral index, the LISA event rate from SMBH mergers would be reduced as well.  \citet{RW05} used EPS merger theory to predict that LISA will have approximately 15 SMBH-merger detections per year at a signal to noise greater than five (they only consider mergers with $M_{\rm BH} \gsim 10^5 \sm $).  These comparisons of EPS and BKH event rates indicate that LISA's event rate may be closer to ten, with all other assumptions held fixed.    

\section{Summary and discussion}
\label{sec:CON}
The EPS merger theory used to predict supermassive-black-hole merger rates is mathematically inconsistent because it contains two merger rates for the same pair of haloes.  When the EPS formalism is used to derive supermassive-black-hole merger rates and the corresponding event rate for LISA, there are two potential results; the EPS predictions are ambiguous.  We have found that mergers between haloes whose masses differ by less than a factor of $10^2$ dominate the SMBH merger rate, even when all mergers between SMBH-containing haloes are included.  The difference between the EPS merger rates for mass ratios in this range is small, so the two merger rates predicted by EPS theory are nearly equal.

The concordance between the two EPS predictions for the SMBH merger rate is an artifact of the relative paucity of haloes with masses larger than $10^{11} \sm $.  It is not an indication that the EPS merger formalism may be trusted to give realistic merger rates.  In addition to its mass-asymmetry, the Lacey--Cole merger rate fails to give the same evolution of the halo population as the time derivative of the Press--Schechter mass function.  Both of these flaws justify the search for a new theory of halo mergers.  \citet*{BKH05} (BKH) inverted the coagulation equation to find merger rates that preserve the Press--Schechter halo mass function for power-law power spectra.  They found that these merger rates differ significantly from the EPS rates for the same power spectrum.     

The limited range of halo masses that contribute to the SMBH merger rate makes it possible to find a power-law power spectrum that accurately fits the mass variance $\sigma(M)$ in this region.  We consider such a power-law approximation with spectral index $n=-2.1$.  Since the EPS merger formula depends only on the values of $\sigma(M)$ for the two halo masses that are merging and the mass of the resulting halo, the power-law approximation accurately describes the result obtained from the \lcdm power spectrum.  The same correspondence cannot be assumed for the BKH merger rates because they are dependent on $\sigma(M)$ at all masses.  

Nevertheless, it is illuminating to compare the SMBH merger rates derived from BKH merger theory to those derived from EPS theory for the same spectral index.  When $n=-2.1$, the BKH merger rates are lower than the corresponding EPS rates for nearly equal-mass halo mergers, which dominate the rate of SMBH mergers.  This discrepancy is a clear demonstration of how the EPS rates fail to solve the coagulation equation and therefore fail to preserve the PS halo mass function.  It also indicates how BKH theory may predict a different SMBH-merger event rate for LISA, since the difference in merger rates results in an equally large difference in event rates.  Comparing the event rates derived from EPS and BKH merger theories for this spectral index indicates that the LISA event-rate predictions that employ EPS merger theory may over-estimate the event rate by thirty percent.
  
Fortunately, the ambiguity carried into the SMBH-merger event-rate predictions for LISA from the uncertainty surrounding halo merger theory does not appear to immediately preclude extracting information regarding reionization or black-hole formation from LISA's observations of SMBH mergers.  \citet{WL03gw} showed that the LISA SMBH-merger event rate with reionization occurring at $z=7$ is about 2.4 times higher than if reionization occurred earlier, at $z=12$.  This difference is larger than the uncertainties in the event rate revealed by our comparisons of BKH and EPS predictions, so it may be possible to constrain the reionization redshift from the LISA SMBH-merger event rate without a definitive theory of halo mergers.  The thirty-percent uncertainty implied by these halo-merger-theory comparisons is also less than the difference in event rates for different SMBH seeding found by \citet{MHN01}.  However, a thirty-percent uncertainty in the SMBH-merger rate will significantly loosen the constraints LISA's observations of SMBH mergers could place on reionization and SMBH formation.  More concerning is the fact that there is no guarantee that the merger rate derived from the merger kernel that satisfies the coagulation equation for a \lcdm universe does not differ from the EPS merger rate by more than thirty percent.    

Clearly, solving the coagulation equation for a \lcdm power spectrum is imperative.  Any application of extended Press--Schechter merger theory to astrophysical phenomena has a flawed foundation and the resulting predictions are unreliable.  Specifically, we have shown that the differences between EPS merger theory and BKH merger theory for power-law power spectra indicate that switching merger theories could significantly alter the LISA SMBH-merger event rate.  This theoretical uncertainty should be resolved before LISA's measurements of SMBH merger rates are used to constrain cosmological models.

We thank Jonathan Pritchard for useful discussions.  ALE is supported by a NSF Graduate Fellowship.  AJB is supported by a Royal Society University Research Fellowship.  This work was supported at Caltech by DoE DE-FG03-92ER40701 and NASA NNG05GF69G.

\end{document}